\documentclass[twocolumn,showpacs,preprintnumbers,amsmath,amssymb,superscriptaddress]{revtex4}
\usepackage{graphicx}
\usepackage{dcolumn}
\usepackage{bm}
\usepackage{soul}
\usepackage{color}
\usepackage{epstopdf}
\usepackage[version=3]{mhchem}
\usepackage{lipsum}
\usepackage[outercaption]{sidecap}
\usepackage{floatrow}
\begin{document}

\title{Repulsive interactions of a lipid membrane with graphene in composite materials}

\author{Anh D. Phan}
\affiliation{Department of Physics, University of South Florida, Tampa 33620, USA}%
\email{anhphan@mail.usf.edu}
\affiliation{Institute of Physics, 10 Daotan, Badinh, Hanoi, Vietnam}%
\author{Trinh X. Hoang}
\affiliation{Institute of Physics, 10 Daotan, Badinh, Hanoi, Vietnam}
\author{The-Long Phan}
\affiliation{Department of Physics, Chungbuk National University, Cheongju 361-763, Korea}
\email{ptlong2512@yahoo.com}
\author{Lilia M. Woods}
\affiliation{Department of Physics, University of South Florida, Tampa 33620, USA}%
\date{\today}

\begin{abstract}
The van der Waals interaction between a lipid membrane and a substrate covered by a graphene sheet is investigated using the Lifshitz theory. The reflection coefficients are obtained for a layered planar system submerged in water. The dielectric response properties of the involved materials are also specified and discussed. Our calculations show that a graphene covered substrate can repel the biological membrane in water. This is attributed to the significant changes in the response properties of the system due to the monolayer graphene. It is also found that the van der Waals interaction is mostly dominated by the presence of graphene, while the role of the particular substrate is secondary.

\end{abstract}

\pacs{}
\maketitle
\section{Introduction}
The cell membrane is an important component of all living organisms \cite{27,28,22}. Its primary role is to provide protection to the cell and sub-cellular structures from their surroundings. The cell membrane is made of a lipid bilayer of amphipathic phospholipid molecules which spontaneously rearrange in a polar solvent such as water \cite{23,24}. The cell membrane's polar surfaces can carry electric charges causing the membrane to be a key element for cellular electrostatics \cite{25,26}. For example, the electrostatic potential difference between the two membrane sides is used by the cell to control the flux of ions through channels in passive transport. Apart from its biological importance, the lipid membrane is an interesting material for technological applications due to its versatility and compelling electrostatic properties.

New nanotechnological devices for chemical, biological, and environmental applications have attracted considerable attention in recent years  \cite{18}. Understanding the forces between cell membranes and substrates is one of the most important steps for designing biological devices. These membrane-substrate interactions have been studied both experimentally \cite{15,16} and theoretically \cite{17,29,30,31}.  Van der Waals and Casimir interactions are of primary importance in biological systems, and the dielectric properties of the cell membrane have been found to be crucial for the description of such forces \cite{5,19}.At the same time, graphene with its remarkable optical, electric, and mechanical properties, has allowed the development of new functional devices. A number of studies have proposed graphene-based devices such as field-effect 
transistors \cite{9,10}, sensors \cite{11,12} and supercapacitors \cite{13,14}. It has also been discovered that bacterial cell membranes can be damaged by sharp-corner graphene sheets \cite{20,39}. Therefore, the use of graphene sheets could be dangerous to the human body. A variety of questions about the processes of graphene-based materials interacting with cell membranes are yet to be answered. In the present work, we present calculations of van der Waals interactions involving a lipid membrane, graphene and different substrates. We show that it is possible to obtain repulsive van der Waals interaction in certain configurations. Our finding can be useful for further advancing of biomedical and biosensing technologies.

\begin{SCfigure*}
\includegraphics[width=12cm]{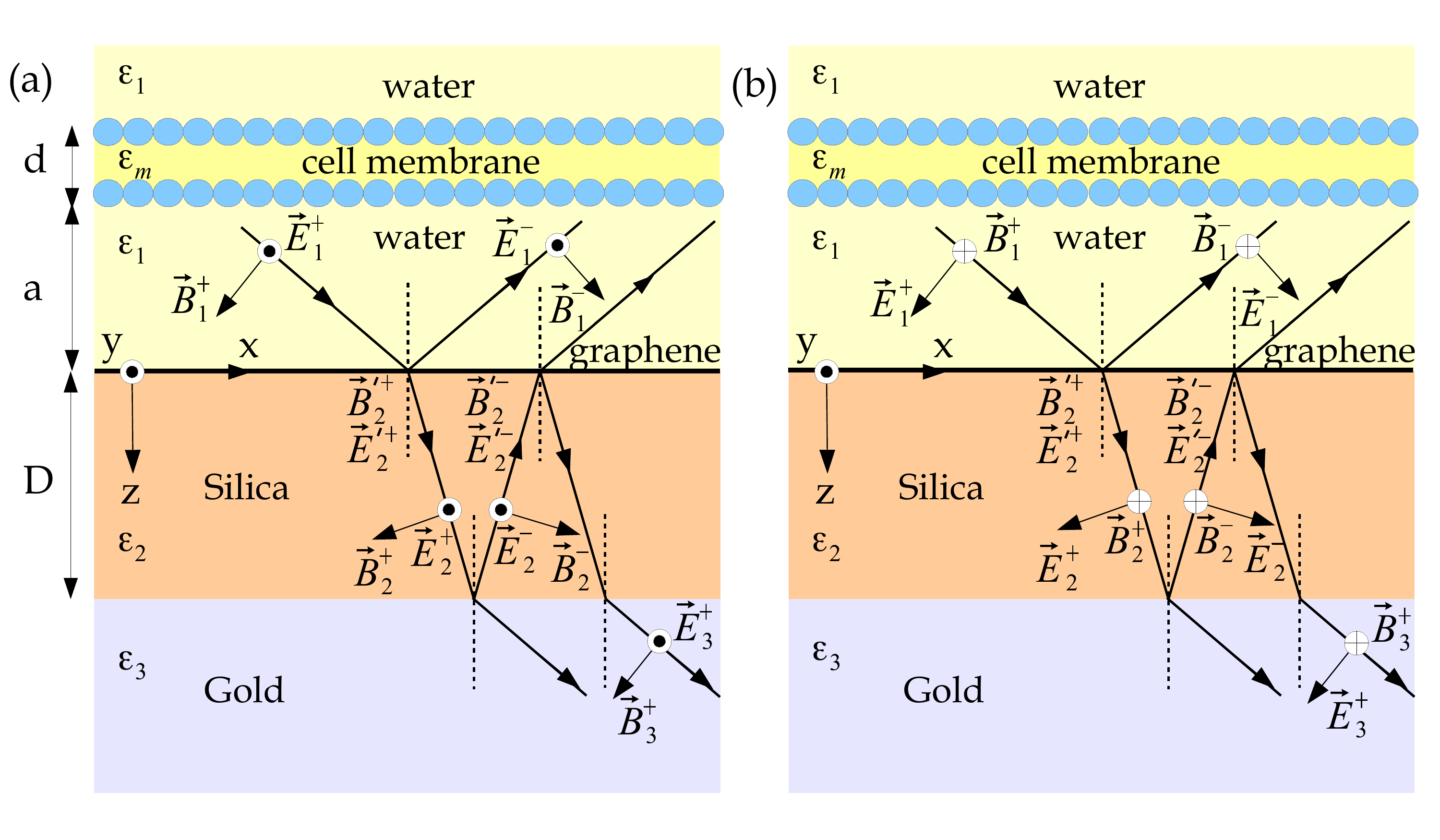}
\hspace{0.2cm}\parbox{0.5cm}{\caption{Schematic picture of a lipid membrane and graphene on the top of a composite substrate including a silica layer with thickness $D$ and a semi-infinite-gold space in water. The dielectric functions of each medium are shown. The boundary condition for the (a) TE and b(TM) modes, necessary for the calculation of the van der Waals force, are also shown.}\label{fig:1}}
\end{SCfigure*}

\section{Theoretical background}

The system under consideration here is shown in Fig. \ref{fig:1}. It consists of a lipid membrane above a graphene covered substrate. The substrate is composed of a silica layer above semi-infinite gold. The surrounding medium is water. The characteristic distances are also specified in Fig. \ref{fig:1}. We are interested in the van der Waals interaction between the lipid membrane separated from the composite substrate (graphene/silica/gold) by a distance $a$. The force per unit area at temperature $T$ is given by \cite{33,34,1,2}
\begin{eqnarray}
&&F(a,T)=\frac{k_{B}T}{\pi}\sum_{l=0}^{\infty}\left(1-\frac{1}{2}\delta_{l0}\right)\int_{0}^{\infty}k_{1z}qdq\times \nonumber \\
&& \left(\frac{R^{TE}_0R^{TE}_1}{e^{2ik_{1z}a}-R^{TE}_0R^{TE}_1}+\dfrac{R^{TM}_0R^{TM}_1}{e^{2ik_{1z}a}-R^{TM}_0R^{TM}_1}\right),
\label{11}
\end{eqnarray}
where $k_B$ is the Boltzmann constant and $\delta_{l0}$ is the Kronecker delta. $R^{TE}_{0,1}\equiv R^{TE}_{0,1}(i\xi_l,q)$ and $R^{TM}_{0,1}\equiv R^{TM}_{0,1}(i\xi_l,q)$ are the reflection coefficients for the transverse electric (TE) and transverse magnetic (TM) field modes, respectively. The boundary conditions for the TE and TM modes are shown in Fig. \ref{fig:1}. Also, $\xi_l=2\pi k_BTl/\hbar$ are the Matsubara frequencies with $\hbar$ being the Plank's constant and $k_{nz}=\sqrt{\varepsilon_n\xi_l^2/c^2+q^2}$ - the magnitude of the wave vector component along the $z$ axis in medium with dielectric function $\varepsilon_n$. $q$ is the wave vector component parallel to the surface and $c$ is the speed of light. According to Figure \ref{fig:1}, the medium between the membrane and the composite substrate is water with $\varepsilon_1$. Finally,  $\omega_l$ are the imaginary frequencies $\omega_l=i\xi_l$.

\subsection{Graphene on substrate}

In order to calculate the van der Waals force, the reflection coefficients are obtained first. These can be found from the Maxwell's equations with charge- and current-free media 

\begin{eqnarray}
\nabla \vec{B} = 0&,& \nabla \vec{E} = 0,\nonumber\\
\nabla \times \vec{B} &=& \mu_0\vec{j}+\frac{\varepsilon}{c^2}\frac{\partial\vec{E}}{\partial t},\nonumber\\
\nabla \times \vec{E} &=& -\frac{\partial\vec{B}}{\partial t},
\label{1}
\end{eqnarray}
where $\mu_0$ is the vacuum permeability and $\varepsilon$ is the dielectric function of the medium. The graphene sheet is located in the interface between medium $1$ and $2$. The current $\vec{j}$ directed by the optical electric field only flows on the surface of graphene. 

The planar boundary conditions of the total electric and magnetic fields crossing from medium 1 to 2 are given by (Fig. \ref{fig:1})
\begin{eqnarray}
E'_1 &=& E'_2,\nonumber\\
H_{2t}-H_{1t}&=&\sigma E'_2,
\label{2}
\end{eqnarray}
where the subscript $t$ denotes for the tangential component of the field and $\sigma$ is the graphene optical conductivity. The graphene conductivity for energies $< 3$ eV has been measured \cite{32} and it is well-described by a two-band Dirac
model \cite{6}. Meanwhile, the higher energy regimes have been challenging for scientists.  We, therefore, focus on the graphene conductivity at a small frequency range because of the fact that the van der Waals interactions are mainly contributed by low-frequency terms (less than 4 eV) \cite{1,6}. In this case, the optical conductivity is expressed as a function of imaginary frequencies $\xi$ by the Kubo formalism \cite{6}
\begin{eqnarray}
\sigma(i\xi)=\frac{2e^2k_{B}T\ln(2)}{\pi\hbar^2\xi}+\frac{e^2\xi}{8\pi k_{B}T}\int_{0}^{\infty}\frac{\tanh(x)dx}{x^2+(\frac{\hbar\xi}{4k_{B}T})^2},
\label{15}
\end{eqnarray}
where $e$ is the charge of an electron. The first term corresponds to the intraband transitions, while the second corresponds to the interband transitions. When $T \rightarrow 0$, $\sigma(\omega)$ takes the universal value $\sigma_0=e^2/4\hbar$ \cite{32,6,50}.

For the TE mode, using Eq.(\ref{1}), one finds that $k_zE = -\omega\mu_0H_t$. This combined with Eq.(\ref{2}) gives (Fig.(\ref{fig:1}):
\begin{eqnarray}
{E}_1^+ + E_1^- &=& {E'}_2^+ + {E'}_2^-,\nonumber\\
\frac{k_{2z}({E'}_2^+ - {E'}_2^-)}{\omega\mu_0}&-&\frac{k_{1z}(E_1^+ - E_1^-)}{\omega\mu_0}+\sigma ( {E'}_2^+ + {E'}_2^-)=0.\nonumber\\
\label{3}
\end{eqnarray}

The expression of the reflection coefficient $R^{TE}_1 = E_1^-/E_1^+ $ on the 1-2 interface is therefore given by

\begin{eqnarray}
\cfrac{\cfrac{k_{1z}-k_{2z}-\mu_0\omega\sigma}{k_{1z}+k_{2z}+\mu_0\omega\sigma}+\cfrac{k_{1z}+k_{2z}-\mu_0\omega\sigma}{k_{1z}+k_{2z}+\mu_0\omega\sigma}\cfrac{{E'}_2^-}{{E'}_2^+}}{1+\cfrac{k_{1z}-k_{2z}+\mu_0\omega\sigma}{k_{1z}+k_{2z}+\mu_0\omega\sigma}\cfrac{{E'}_2^-}{{E'}_2^+}}.
\label{4}
\end{eqnarray}

In medium $2$, the relation between the reflection coefficients on the 2-1 interface and the 2-3 interface is expressed by 
\begin{eqnarray}
R^{TE}_2=\frac{{E}_2^-}{{E}_2^+}=\frac{{E'}_2^-}{{E'}_2^+}e^{-2ik_{2z}D},
\label{5}
\end{eqnarray}
where $D$ is the thickness of the layer $2$. If layer 3 is semi-infinite, the reflection coefficient $R^{TE}_2$ can be obtained utilizing the method used for calculating $R^{TE}_1$
\begin{eqnarray}
R^{TE}_2=\frac{{E}_2^-}{{E}_2^+}=\frac{k_{2z}-k_{3z}}{k_{2z}+k_{3z}}.
\label{6}
\end{eqnarray}

For the TM mode, the magnetic fields in all media are along the $y$ direction and $k_zH = -\omega\varepsilon\varepsilon_0E_t$. Eq.(\ref{2}) reads
\begin{eqnarray}
{E'}_{1t}^- - {E'}_{1t}^+ &=& {E'}_{2t}^- - {E'}_{2t}^+,\nonumber\\
-({H'}_2^+ + {H'}_2^-) + ({H'}_1^+ + {H'}_1^-) &=& \sigma({E'}_{2t}^- - {E'}_{2t}^+).
\label{7}
\end{eqnarray}

The reflection coefficient $R_1^{TM}=H_1^-/H_1^+$ can be calculated by
\begin{eqnarray}
\cfrac{\left(1+\cfrac{\sigma k_{2z}}{\omega\varepsilon_2\varepsilon_0} - \cfrac{\varepsilon_1k_{2z}}{\varepsilon_2k_{1z}}\right)+\left(1-\cfrac{\sigma k_{2z}}{\omega\varepsilon_2\varepsilon_0} + \cfrac{\varepsilon_1k_{2z}}{\varepsilon_2k_{1z}}\right)\cfrac{{H'}_2^-}{{H'}_2^+}}{\left(1+\cfrac{\sigma k_{2z}}{\omega\varepsilon_2\varepsilon_0} + \cfrac{\varepsilon_1k_{2z}}{\varepsilon_2k_{1z}}\right)+\left(1-\cfrac{\sigma k_{2z}}{\omega\varepsilon_2\varepsilon_0} - \cfrac{\varepsilon_1k_{2z}}{\varepsilon_2k_{1z}}\right)\cfrac{{H'}_2^-}{{H'}_2^+}},\nonumber\\
\label{8}
\end{eqnarray}
where
\begin{eqnarray}
\cfrac{{H'}_2^-}{{H'}_2^+}=\cfrac{{H}_2^-}{{H}_2^+}e^{2ik_{2z}D}.
\label{9}
\end{eqnarray}

The reflection coefficient $TM$ of the semi-infinite medium 3 is found to be
\begin{eqnarray}
R^{TM}_2=\frac{{H}_2^-}{{H}_2^+}=\frac{\varepsilon_3 k_{2z}-\varepsilon_2 k_{3z}}{\varepsilon_3 k_{2z}+\varepsilon_2 k_{3z}}.
\label{10}
\end{eqnarray}

The usual solvent for lipid membranes is an aqueous solution whose dielectric response is close to pure water. The dielectric function of this liquid versus imaginary frequencies are modelled by \cite{51}
\begin{eqnarray}
\varepsilon(i\xi)=1+\sum_{k=1} \frac{d_k}{1+\xi\tau_k} + \sum_{s=1}\frac{f_s}{\omega_s^2+\xi^2+\xi g_s},
\label{13}
\end{eqnarray}
where the first sum is responsible for Debye relaxation and the second sum is described by the summation of the damped harmonic
oscillators. $d_k$ and $\tau_k$ are the peak height and relaxation time, respectively. $\omega_s$ is the resonant frequency, $g_s$ is the damping parameter, and $f_s$ is the oscillatory strength.

In our calculation, the thin film $\varepsilon_2$ is $\ce{ SiO_2}$. The dielectric function of silica is represented the Lorentz oscillator model 
\begin{eqnarray}
\varepsilon(i\xi)=1+\sum_i \frac{C_i\omega_i^2}{\xi^2 + \omega_i^2},
\label{12}
\end{eqnarray}
in which $\omega_i$ is the resonant frequency and $C_i$ is the oscillatory strength at $\omega_i$. All parameters of the oscillator model in our paper taken from Ref.\cite{3} were fitted with data measured in a wide range of frequencies. The accuracy of this model has been examined in measurements of the Casimir force \cite{3}.

The dielectric function of gold along imaginary frequencies is modelled by the Drude model \cite{2,3}
\begin{eqnarray}
\varepsilon(i\xi)=1+\frac{\omega_{Au,p}^2}{\xi(\xi + \gamma_{Au})},
\end{eqnarray}
where $\omega_{Au,p} = 9$ eV and $\gamma_{Au} = 0.035$ eV are the plasma frequency and the damping parameter of gold, respectively. 
\subsection{Lipid membrane}
To interpret the dielectric function of a phospholipid membrane, we consider the motion of a bound electron in the membrane governed by the external field of light $\vec{E} = \vec{E}_0e^{-i\omega t}$. The equation of motion can be expressed by \cite{8}
\begin{eqnarray}
m\frac{d^2\vec{r}}{dt^2}=e\vec{E}-m\omega_0^2\vec{r}-m\gamma\frac{d\vec{r}}{dt}.
\label{16}
\end{eqnarray}
where $m$ is the mass of electron, $e$ is the electron charge, $\omega_0$ is the characteristic frequency, $\gamma$ is the damping parameter describing the friction between the electron and environment during motion.

After straightforwardly solving Eq.(\ref{16}), the susceptibility can be found
\begin{eqnarray}
\chi = \frac{\omega_p^2}{\omega_0^2-\omega^2-i\gamma\omega},
\label{17}
\end{eqnarray}
where $\omega_p = Ne^2/m\varepsilon_0$ and $N$ is the bound electron density.

Comparing the above result with the dielectric function of the membrane given by \cite{4,5}, the oscillator model is capable of describing the dielectric function of the membrane $\varepsilon_m(i\xi)$. Experimental data shows that a single oscillator captures the dielectric response of a lipid membrane in the low frequency regime \cite{4,5}.

\begin{eqnarray}
\varepsilon_m(i\xi)=1+\frac{[\varepsilon_m(0)-1]\omega_{uv}^2}{\xi^2 + \omega_{uv}^2},
\label{13}
\end{eqnarray}
where $\varepsilon_m(0)=2$ is the dielectric constant of the membrane at zero frequency and $\omega_{uv} \approx 6.6$ eV is the characteristic frequency \cite{4,5}. From Eq.(\ref{17}) and Eq.(\ref{13}), we obtain $\omega_0 = \omega_{uv}$, $\gamma = 0$ and $\omega_p^2=\left[\varepsilon_m(0)-1\right]\omega_{uv}^2$. 

The reflection coefficient for the membrane 
$R_0^{TE,TM}$ are found in the same way as for $R_1^{TE,TM}$, however, here $\sigma = 0$.
\begin{eqnarray}
R^{TE}_0=\frac{k_{1z}-k_{mz}}{k_{1z}+k_{mz}}\cfrac{1-e^{2ik_{mz}d}}{1-\left(\cfrac{k_{1z}-k_{mz}}{k_{1z}+k_{mz}}\right)^2e^{2ik_{mz}d}},\nonumber\\
R^{TM}_0=\frac{\varepsilon_m k_{1z}-\varepsilon_1 k_{mz}}{\varepsilon_m k_{1z}+\varepsilon_1 k_{mz}}\cfrac{1-e^{2ik_{mz}d}}{1-\left(\cfrac{\varepsilon_m k_{1z}-\varepsilon_1 k_{mz}}{\varepsilon_m k_{1z}+\varepsilon_1 k_{mz}}\right)^2e^{2ik_{mz}d}}.\nonumber\\
\label{14}
\end{eqnarray}

\section{Numerical results and discussions}
We consider the case when there is no substrate under the graphene sheet, thus the system consists of a lipid membrane and a graphene sheet in water.  The reflection coefficient of the bottom substrate is given by
\begin{eqnarray}
R^{TE}_1(i\xi_l,q)&=&\frac{-\mu_0\xi\sigma(i\xi_l)}{2k_{1z}+\mu_0\xi_l\sigma(i\xi_l)},\nonumber\\
R^{TM}_1(i\xi_l,q)&=&\frac{\sigma(i\xi_l) k_{1z}}{2\varepsilon_0\varepsilon_1(i\xi_l)\xi +\sigma(i\xi_l) k_{1z}}.
\label{18}
\end{eqnarray}

\begin{figure}[htp]
\includegraphics[width=9cm]{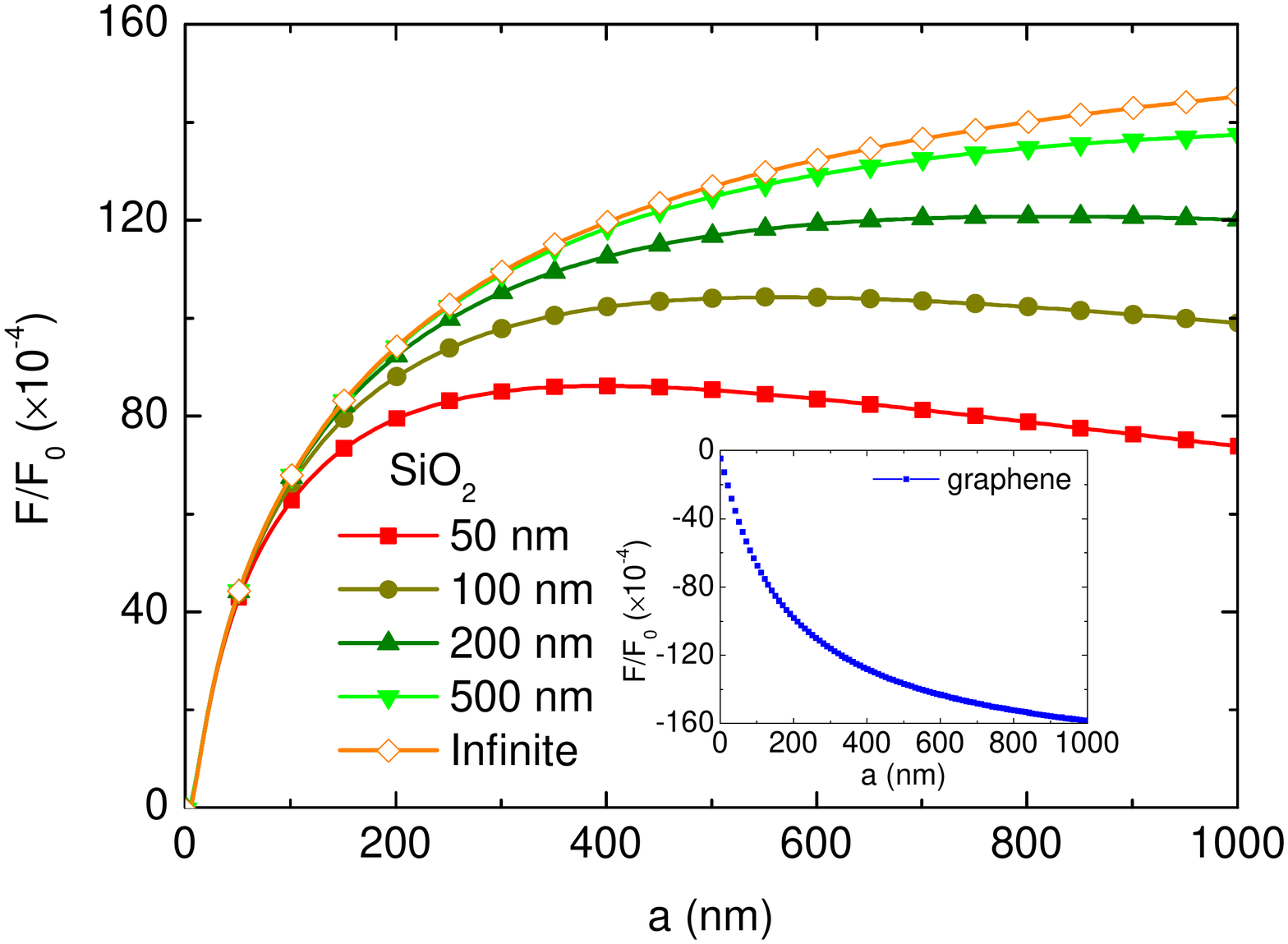}
\caption{\label{fig:2}(Color online) The van der Waals interactions between a lipid membrane and $\ce{SiO_2}$ with different thicknesses in the absence of the gold layer. The inset figure shows graphene without a substrate below repelling the biological membrane. Here $F_0=-\pi^2\hbar c/240a^4$ is the Casimir force between two ideal metal plate.}
\end{figure}
wherefrom it follows that $R^{TE}_1(i\xi,q) < 0$ and $R^{TM}_1(i\xi,q) > 0$. Similarly, since  $\varepsilon_1(i\xi_l) > \varepsilon_m(i\xi_l)$, we have $R_{0}^{TE}(i\xi_l,q) > 0$ and $R_{0}^{TM}(i\xi_l,q) < 0$. The fact that the reflection coefficients have different signs changes also the signs of the van der Waals interaction which turns repulsive, as shown in the inset of Fig.\ref{fig:2}. 

When the lipid membrane interacts with a silica film with a finite thickness, $\sigma(i\xi_l) = 0$, $\varepsilon_2(i\xi_l)=\varepsilon_{\ce{SiO_2}}(i\xi_l)$ and $\varepsilon_3(i\xi_l) = \varepsilon_1(i\xi_l)$. Figure \ref{fig:2} shows attractive van der Waals interaction between the lipid membrane and thin films with various thicknesses at $T = 300$ $K$. At very small separations $a \le 4.5$ $nm$, we find that the force is repulsive, indicating that the cell membrane cannot adhere to the silicon dioxide surface. This change of sign in the interaction may indicate that a stable equilibrium of the membrane above a silica film may be observed. We also note that since the van der Waals force is not affected by the $SiO_2$ thickness for $a \le 100$ $nm$. 

The similar change of the sign of the van der Waals interaction has been observed in other systems but for different reasons \cite{3,2}. The critical spacing, where the van der Waals interaction changes sign, observed for the system discussed is very close to the case of water-one-ice \cite{35}. In both cases there is thus an equilibrium distance set only by the van der Waals interactions.

The results in Fig. \ref{fig:2} also suggest that the repulsive-attractive transition can be used for designing graphene-based devices trapping lipid membrane. We also find that the dispersion interaction is temperature independent. At short distances $\hbar c/(2\pi k_BTd) \gg 1$, both experimental \cite{3} and our theoretical calculations show that the van der Waals interactions at $T = 300$ and $0$ $K$ can be calculated by the same expression. It was reported that the sum in Eq.(\ref{11}) can be can be transformed into an integral \cite{6} at such range. For longer ranges, the contribution of the film thickness becomes much more considerable due to thermal fluctuation effects. The increase of $D$ increases the magnitude of the attraction. The $F/F_0$ lines corresponding to the silica with $D=500$ $nm$ and $D = \infty$ are identical when $a \le 500$ $nm$. If graphene is coated on the interface of medium 2 and 3, the force remains unchanged due to the significant screening from 
the silicon dioxide film. These results suggest that the van der Waals force strongly depends on the surfaces of objects. 

\begin{figure}[htp]
\includegraphics[width=9cm]{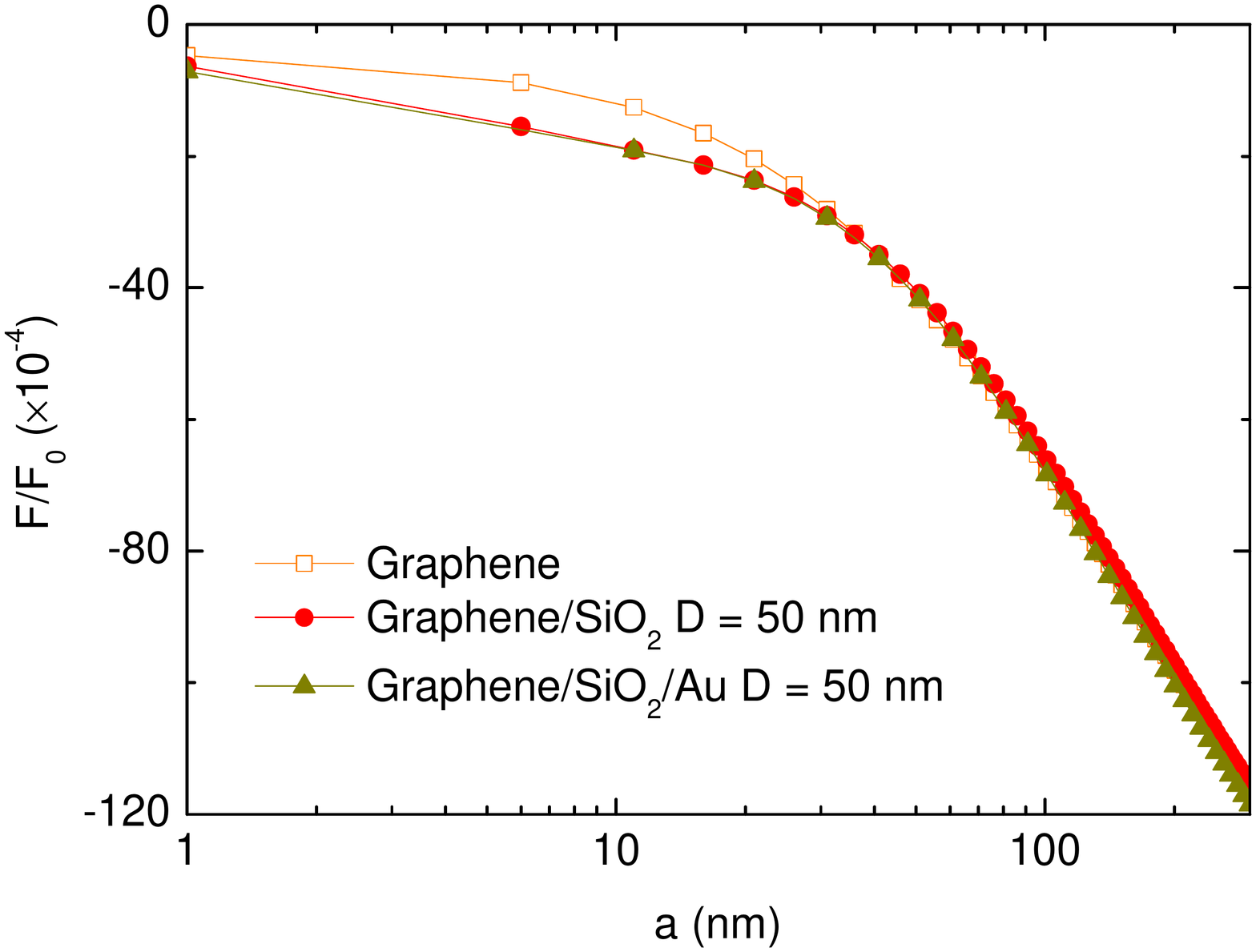}
\caption{\label{fig:3}(Color online) The ratio $F/F_0$ of a lipid membrane with graphene, graphene on \ce{SiO_2} and graphene on the two-layer system including \ce{SiO_2} thin film and infinite gold substrate at $T = 300$ $K$.}
\end{figure}

Next, we consider the case when the silica film is covered by the graphene sheet. As can be seen from Fig.\ref{fig:3}, silicon dioxide is almost invisible at the distance $a \ge 30$ nm, suggesting that graphene has a major effect on the repulsive van der Waals force. Inserting semi-infinite gold under the $SiO_2$ film does not introduce significant changes in the force. At distances $a \le 30$ nm, however, the membrane attraction towards the graphene covered substrate is stronger as compared to the case when only graphene is present. The fact that $F/F_0$ differs little for the different substrates at $a> 80$ nm in  Fig.\ref{fig:3} shows that the graphene is dominating the van der Waals interaction.

\begin{figure}[htp]
\includegraphics[width=9cm]{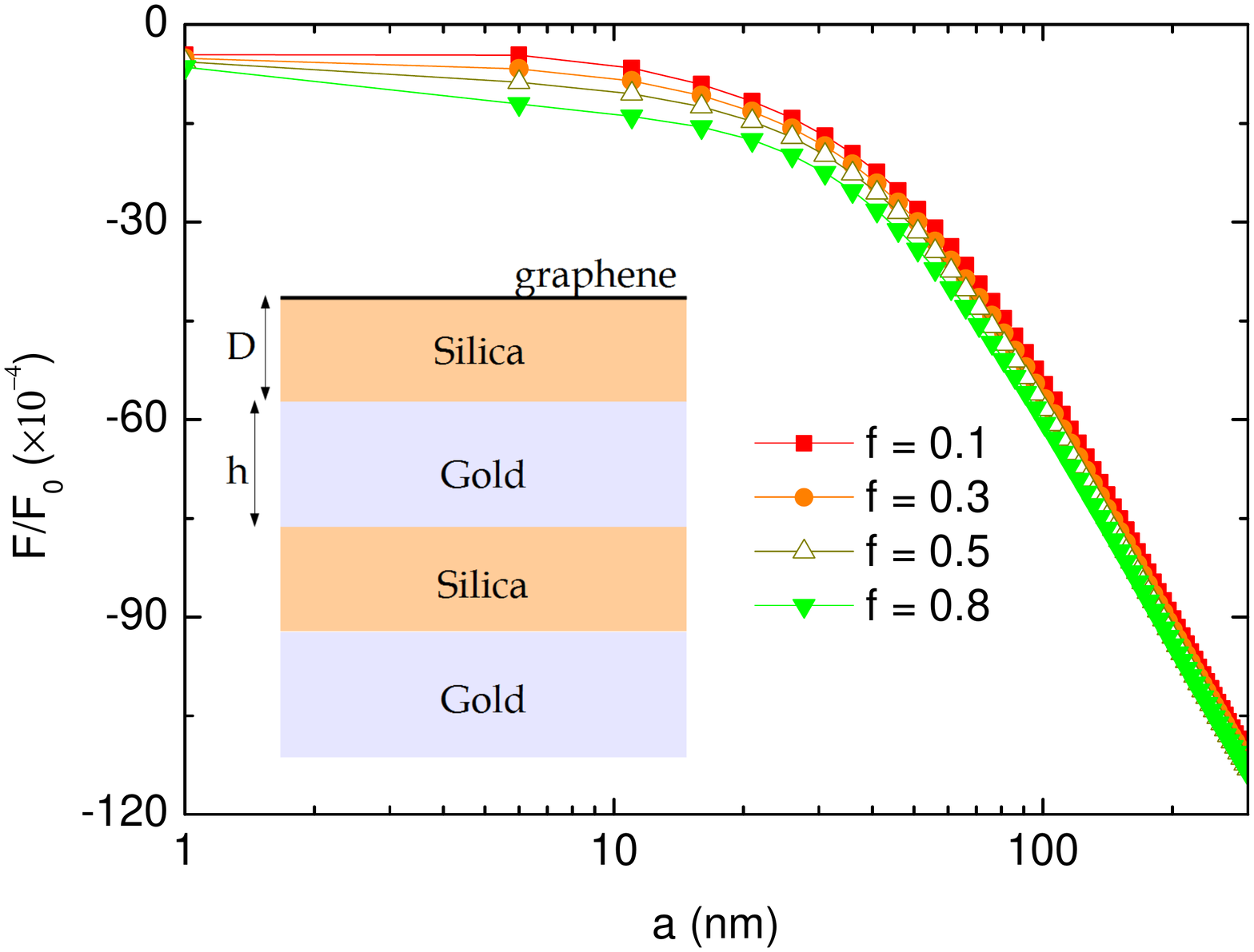}
\caption{\label{fig:4}(Color online) The van der Waals force of a lipid membrane and a graphene sheet on a superlattice substrate composed of alternating silica and gold layers at $T = 300$ $K$ for different fractions $f$. The insert shows the graphene covered composite substrate with corresponding layer thicknesses.}
\end{figure}

Finally, we consider the situation of a superlattice substrate with periodic arrangement of $SiO_2/Au$ alternating layers with different thicknesses (insert of  Fig.\ref{fig:4}) covered with a graphene sheet. One can calculate the permittivities of this system by means of an anisotropic effective dielectric function:
\begin{eqnarray}
\varepsilon_{\perp}&=&f\varepsilon_{\ce{SiO_2}} + \left(1-f \right)\varepsilon_{\ce{Au}},\nonumber\\
\varepsilon_{\parallel}&=&\frac{\varepsilon_{\ce{SiO_2}}\varepsilon_{\ce{Au}}}{f\varepsilon_{\ce{Au}}+\left(1-f \right)\varepsilon_{\ce{SiO_2}}},
\label{19}
\end{eqnarray}
where $f = D/(D+h)$ is the fraction of silica in the composite material. The reflection coefficients of the bottom object are then given by

\begin{eqnarray}
R^{TE}_1&=&\cfrac{\sqrt{\varepsilon_1\cfrac{\xi_l^2}{c^2}+q^2}-\sqrt{\varepsilon_{\parallel}\cfrac{\xi_l^2}{c^2}+q^2}-\sigma\mu_0\xi_l}{\sqrt{\varepsilon_1\cfrac{\xi_l^2}{c^2}+q^2}+\sqrt{\varepsilon_{\parallel}\cfrac{\xi_l^2}{c^2}+q^2}+\sigma\mu_0\xi_l},\nonumber\\
R^{TM}_1&=&\cfrac{\varepsilon_{\parallel}/\sqrt{\varepsilon_{\parallel}\cfrac{\xi_l^2}{c^2}+q^2\cfrac{\varepsilon_{\parallel}}{\varepsilon_{\perp}}}-\varepsilon_{1}/\sqrt{\varepsilon_1\cfrac{\xi_l^2}{c^2}+q^2}+\cfrac{\sigma}{\varepsilon_0\xi_l}}{\varepsilon_{\parallel}/\sqrt{\varepsilon_{\parallel}\cfrac{\xi_l^2}{c^2}+q^2\cfrac{\varepsilon_{\parallel}}{\varepsilon_{\perp}}}+\varepsilon_{1}/\sqrt{\varepsilon_1\cfrac{\xi_l^2}{c^2}+q^2}+\cfrac{\sigma}{\varepsilon_0\xi_l}}.\nonumber\\
\label{20}
\end{eqnarray}

Figure \ref{fig:4} illustrates the van der Waals interaction between the lipid membrane and graphene on the periodic substrate. One finds that the force has a similar behavior and magnitude as in the configuration displayed in  Fig.\ref{fig:3}. The more significant changes are found in the short distance separation for the different $f$ factors. These results emphasize again the dominating effect of the graphene on the surface of the substrate on the interaction. It also appears that the first $SiO_2$ layer affects the magnitude in this distance separation range, similar to the results shown in  Fig.\ref{fig:3}, while the rest of the superlattice has practically no effect on the interaction. 

The exact solution for the reflection coefficients of multilayered systems found by the transfer matrix method in Ref. \cite{36,37,38} allows us to precisely calculate the van der Waals interactions for the system in Fig.\ref{fig:4}. The results are consistent with all expression of the reflection coefficients in our section II.

In the case of the periodically multilayered system in Fig.\ref{fig:4} with parameter $D=50$ $nm$ and $h = 25$ $nm$, remarkably, the van der Waals forces are similar to what we presented in Fig.(\ref{fig:3}) for a two-layer system. Note, however, that $h$ should be greater than 25 $nm$ to avoid the skin depth effects \cite{2}. Increasing $h$ does not modify the van der Waals force because the gold thickness 25 $nm$ can be considered already as a gold semi-infinite space in the Lifshitz calculations. Prolifering the number of layers therefore does not introduce any fundamentally new features if compared with a simple two-layer system. 
\section{Conclusions}
We have studied the van der Waals interaction of a phospholipid membrane with a multilayer object involving graphene in water. A silica thin film repels graphene at distances $a \le 4.5$ $nm$ and the interaction becomes attractive at larger distances. The position of the transition is independent of the silica thickness due to the fact that the stable separation is much smaller than the considered thickness. This feature can be exploited to create devices trapping lipid membranes. The van der Waals force in such system can be entirely repulsive by using a graphene sheet on the silica thin film. At $a \ge 30$ $nm$, the dispersion forces are dominated by the graphene, the silica layer only affects the force at shorter separation distances. When putting graphene on a composite material of two layers: silica and gold, the influence of the metal layer on the interaction of the graphene with the membrane begin to diverge from the van der Waals interaction in the case of the absence of the gold layer at $a \ge 80$ $nm$. This finding suggests that the effect of the layer of gold on this van der Waals force can be ignored due to the optical response of graphene on the surface and the screening of the silica thin film. Our result shows that the model of an effective dielectric function for multi-layer systems is not valid for calculations of the van der Waals interactions.
\begin{acknowledgments}
This work was supported by the Nafosted Grant No. 103.01-2013.16. L.M.W. also acknowledges financial support from the US Department of Energy under Contract DE-FG02-06ER46297.
\end{acknowledgments}

\end{document}